\documentclass[11pt]{article}  
\usepackage{menuproc}
\usepackage{cite}
\usepackage{epsfig}
\def\be{\begin{eqnarray}}
\def\ee{\end{eqnarray}}
\def\vA{{\mathbf  A}}

\def\He#1{{}^{#1}\mbox{He}}
\def\nlo#1{\mbox{N$^{#1}$LO}}
\def\Sunit{\mbox{$10^{-20}$ keV-b}}
\def\dR{{\hat d^R}}
\begin{document}
%
%
%
\titlematter{The Solar $hep$ Processes in Effective Field Theory}%
{Tae-Sun Park$^a$, Kuniharu Kubodera$^a$, Dong-Pil Min$^b$,
and Mannque Rho$^{b,c,d}$}%
{${}^a$ Department of Physics and Astronomy,
University of South Carolina, Columbia, SC 29208, USA\\
 ${}^b$ School of Physics and Center for Theoretical Physics,
Seoul National University, Seoul 151-742, Korea\\
${}^c$ Service de Physique Th\'{e}orique, CEA  Saclay,
\it
91191 Gif-sur-Yvette Cedex, France\\
${}^d$ Institute of Physics and Applied Physics, Yonsei
University, Seoul 120-749, Korea}%
{
By combining effective field theory with
the standard nuclear physics approach (SNPA)
we obtain a high-precision estimate of
the $S$ factor for the solar $hep$ process.
The accurate wave functions
available in SNPA are used to evaluate
the nuclear matrix elements for the transition operators
that result from chiral perturbation theory (ChPT).
All the contributions up to \nlo3 in ChPT 
are included.
The resulting parameter-free, 
error-controlled prediction is:
$S(hep)=(8.6 \pm 1.3 )\times \Sunit$.
}
%

\noindent 
This brief report is based
on the results of work
done in collaboration with L.E. Marcucci, 
R. Schiavilla, M. Viviani, A. Kievsky 
and S. Rosati~\cite{coll,inpc}. 
A detailed exposition
of the basic ideas underlying our approach 
can be found in \cite{strategy}.

\section{Introduction}

The $hep$ process in the Sun
\be
\He3+p\rightarrow \He4 + e^+ + \nu\,.
\ee
produces the highest energy
solar neutrinos, $E_\nu^{\rm max}(hep)$$\simeq$20~MeV.
The $hep$ neutrinos therefore 
can influence the interpretation
of the results of a recent Super-Kamiokande experiment
that have raised many important issues
concerning the solar neutrino problem and neutrino
oscillations~\cite{controversy,monderen}.
It is to be noted that
the reliable estimation of the $hep$ cross section,
indispensable for addressing these issues, 
is a long-standing challenge for nuclear physics~\cite{challenge}. 
This is mainly because
the leading one-body contributions are highly suppressed
and furthermore the chiral filter mechanism 
-- which allows us to accurately estimate 
many-body corrections --
is ineffective for this process.
For a detailed discussion, see Ref.\cite{coll,inpc}.

The objective of our present work is 
to obtain a significantly improved estimate
of the $hep$ rate using effective field theory (EFT). 
To this end, we adopt a strategy
that exploits the known merits of 
the standard nuclear physics approach (SNPA)
and heavy-baryon chiral perturbation theory 
(HBChPT) simultaneously. 
HBChPT, a well established low-energy EFT of QCD, 
is used to calculate the transition operators; 
{\it all} the operators up to
next-to-next-to-next-to-leading order (\nlo3) 
will be considered. 
The evaluation of the corresponding nuclear matrix elements 
requires highly accurate nuclear wave functions. 
Although it is, at least in principle,
possible to derive from HBChPT
nuclear wave functions to a specified chiral order, 
we choose not to do so. 
Instead, we use realistic wave functions 
obtained in SNPA. 
The power of the proposed scheme 
is the ability to correlate the beta
decay processes in the $A=2, 3, 4$ nuclei.
Thus, as explained in more detail below, 
if the single unknown constant in our EFT
is fixed using one of these processes, 
then we can make totally parameter-free predictions
for the remaining processes.

\section{Theory and Results}

The $hep$ process is dominated 
by the Gamow-Teller (GT) transition,
and hence the reliability of the $hep$ rate estimate
is essentially governed by precision
with which one can calculate the GT amplitude.
According to the chiral counting rule \cite{weinberg}, 
the leading order contributions are due to the well-known
one-body (1B) operators,
and the first corrections arise from \nlo3 two-body (2B)
currents that are suppressed by $(Q/\Lambda_\chi)^3$
compared to the 1B.
Here $Q$ stands for the typical three-momentum scale
and/or the pion mass, and $\Lambda_\chi \sim 1\ \mbox{GeV}$
is the chiral scale.
As stated, we consider here
{\it all} the contributions up to \nlo3.
It is worth emphasizing 
that three-body currents, which are \nlo4,
can be legitimately ignored 
in our \nlo3 calculation.

The \nlo3 2B currents consist of
the one-pion-exchange (OPE) 
and nucleon-nucleon contact-term (CT) parts,
$\vA_{{\rm 2B}} =
\vA_{{\rm 2B}}(\mbox{OPE}) +
\vA_{{\rm 2B}}(\mbox{CT}).$
With the low-energy constants fixed 
from $\pi N$ data~\cite{csTREE},
the OPE part is completely determined.
On the other hand, the CT part contains one parameter, $\dR$,
whose direct evaluation from QCD is not available at present.
Fortunately, it turns out that 
tritium $\beta$-decay, $\mu$-d capture  
and $\nu$--$d$ scattering are sensitive
to the same parameter, $\dR$,
and that they do not depend on any additional parameters 
up to \nlo3.
Thus, any of these processes 
can give the renormalization condition
to fix the value of $\dR$.
Here we choose to use the tritium $\beta$-decay
rate, $\Gamma_\beta$, which is accurately known
experimentally~\cite{schiavilla}. 
Once $\dR$ is fixed, our
calculation involves no unknown parameters.

We calculate the matrix elements of the transition operators
with state-of-the-art
realistic nuclear wave functions.
We employ the correlated-hyperspherical-harmonics (CHH)
wave functions, obtained with the
Argonne $v_{18}$ (Av18) potential
(supplemented with the Urbana-IX three-nucleon
potential for the $A\ge 3$ nuclei) \cite{MSVKRB}.
To control short-range physics in a consistent manner,
we apply the regulator
\be S_\Lambda(q^2) =
\exp\left(-\frac{q^2}{2\Lambda^2}\right). \label{regulator} \ee
to all the nuclear systems in question.
The cutoff parameter $\Lambda$ characterizes the energy-momentum
scale of our EFT.

The value of $\dR$ determined
from the experimental value of $\Gamma_\beta$ is
$\dR=(1.00 \pm 0.07,\ 1.78 \pm 0.08,\ 3.90 \pm 0.10)$.
Here and hereafter, parenthesized three numbers
correspond to the three choices of $\Lambda$, 
$\Lambda$ = 500, 600 and 800 MeV,
in this order.
To see the role of the $\dR$-term,
it is informative to look at $\delta_{\rm 2B}$,
the ratio of the 2B matrix element to that of 1B.
With only the OPE part taken into account, we have
$\delta_{\rm 2B}^{\rm OPE}= (-1.1,\ -1.5,\ -2.0)$.
The inclusion of the $\dR$ term leads to
$\delta_{\rm 2B}^{\rm N3LO}= 
\delta_{\rm 2B}^{\rm OPE+CT}= (-0.60,\ -0.64,\ -0.73)$.
Thus the $\dR$-term drastically reduces the $\Lambda$-dependence
of the 2B contribution;
we see only $\sim$10 \% variation for the entire range
of $\Lambda$ under study.
The $\Lambda$-dependence in the total GT amplitude becomes
more pronounced due to a strong cancellation between the 1B
and 2B terms, but this amplified $\Lambda$-dependence still
remains within acceptable levels.

In addition to the ${}^3S_1$ contributions governed by the
GT amplitude, there are also tiny ${}^1S_0$ and sizable
$P$-wave contributions.
The latter have little $\Lambda$-dependence ($<2$ \%),
and responsible for about one-third of the total $S$ factor.
Adding all the contributions,
the $S$-factor at threshold reads
\be
S(hep)=(8.6 \pm 1.3 )\times \Sunit\,,\label{prediction}
\ee
where the ``error"
spans the range of the $\Lambda$-dependence
for $\Lambda$ = 500--800 MeV.
This result is to be compared to the latest SNPA estimate
in Ref.~\cite{MSVKRB}: $S=9.64 \times \Sunit$.

\section{Discussion}
By determining the only parameter of the theory, $\dR$, 
from the experimental data on triton beta decay, 
we have succeeded in estimating the $hep$ S-factor
in a parameter-free and error-controlled manner.
Our HBChPT calculation (up to \nlo3) gives
a much more accurate estimate than hitherto available.
The EFT results turn out to give support 
to those obtained in the latest SNPA 
calculation \cite{MSVKRB}.

To decrease the uncertainty 
in Eq.(\ref{prediction}), we need to reduce the
$\Lambda$-dependence in the two-body GT term. According to a
general {\it tenet} of EFT, the $\Lambda$-dependence should
diminish when higher order terms are included.
A preliminary study indicates 
that it is indeed possible to
reduce the $\Lambda$-dependence significantly 
by including \nlo4 corrections.
In this connection, we remark that
the $hen$ process, $\He3+n\rightarrow \He4 + \gamma$,
seems very interesting to look into.
The $hen$ process shares many features with the $hep$ 
including the suppression of the 1B matrix element 
and the structure of the many-body currents.
Accurate experimental data 
are available for the $hen$ cross sections, 
but so far no theoretical calculations have succeeded 
in explaining the data quantitatively.
Thus applying the same EFT technique 
to the $hen$ process \cite{song} is very interesting,
and that will also provide a useful check
of the formalism employed in our estimation of $S(hep)$.

\end{document}